\documentclass[11pt]{article}
\usepackage{epsfig}
\begin{document}

\thispagestyle{empty}
\begin{flushright}
MIT-CTP-2935 \\
HLRZ2000\_1 \\
hep-lat/0002002 
\end{flushright}
\begin{center}
\vspace*{5mm}
{\LARGE New approximate solutions of the Ginsparg-Wilson 
\vskip2mm 
equation - tests in 2-d}
\vskip12mm
{\bf Christof Gattringer}
\vskip1mm
Massachusetts Institute of Technology \\
Center for Theoretical Physics \\
77 Massachusetts Avenue, Cambridge MA 02139 USA
\vskip6mm
{\bf Ivan Hip}
\vskip1mm
NIC - John von Neumann Institute for Computing \\
FZ-J\"ulich, 52425 J\"ulich, Germany.
\vskip14mm
\begin{abstract}
A new method for finding approximate solutions of 
the Ginsparg-Wilson equation is tested in 2-d.
The Dirac operator is first constructed and then used in 
a dynamical simulation of the 2-flavor Schwinger model. 
We find a very small mass of the 
$\pi$-particle implying almost chirally symmetric fermions.
The generalization of our method to 4-d is straightforward.
\end{abstract}
\vskip5mm
{\sl To appear in Physics Letters B.}
\end{center}
\newpage
\setcounter{page}{1}
\noindent
{\bf Introduction}
\vskip1mm
\noindent
The last two years have seen a tremendous increase in our understanding of
chiral fermions on the lattice (see \cite{Ni99,Ne99,Lu99} for reviews of
recent developments). It has 
been understood  that the anti-commutator characterizing the chiral symmetry
in the continuum has to be replaced by the Ginsparg-Wilson equation 
\cite{GiWi} 
\begin{equation}
\gamma_5 D \; + \; D \gamma_5 \; = \; D \gamma_5 D \; ,
\label{giwi}
\end{equation}
for the lattice Dirac operator $D$. Based on this relation, chiral symmetry 
can be defined on the lattice and at least for abelian fields chiral gauge
theories can be formulated on the lattice \cite{Lu99}. 

Bringing in the harvest from these developments based on the Ginsparg-Wilson 
equation now hinges on our ability to find solutions for this equation. 
So far two types of solutions are known: The perfect action approach to 
field theories on the lattice (see \cite{perfect} for an extensive review)
produces the fixed point Dirac operator 
\cite{fixpd} which
is a solution of (\ref{giwi}). Although conceptually very beautiful - in 
addition to chirality on the lattice the fixed point operator has other 
nice properties such as e.g.~perfect scaling - constructing the perfect 
Dirac operator is technically very demanding and so far it has been 
implemented only
in two dimensions \cite{fix2d}-\cite{fix2db}. 
The other known solution to (\ref{giwi}) is 
Neuberger's Dirac operator \cite{Neub1} which goes back to the overlap
approach to chiral fermions on the lattice \cite{overlap}. Neuberger's 
construction provides a beautiful and surprisingly simple projection of any 
decent lattice Dirac operator onto a solution of the Ginsparg-Wilson 
equation. However, the numerical evaluation of the inverse
square root necessary for Neuberger's projection seems to be 
a challenging and computationally expensive problem. We should remark, 
that besides perfect actions and the overlap formulation also the domain
wall approach which uses an additinal dimension gives rise to fermions 
with exact chiral symmetry \cite{domainwall}. 

Here we present a new approach to finding 
approximate solutions of the Ginsparg-Wilson equation,
in particular we test the new method in 2-d. In 4 dimensions the 
approach proceeds in exactly the same way and a detailed description
of the new method in 4-d will be presented elsewhere \cite{future}. 

The central idea is to expand the most general Dirac operator $D$ on the
lattice in a suitably chosen basis. Inserting this expression into
the Ginsparg-Wilson equation renders a system of quadratic
equations for the expansion coefficients. The basis for $D$ can be chosen 
in such a way, that there is a natural cut-off for the expansion - the 
length of the paths allowed in the basis elements - and the remaining finite
system of quadratic equations can be solved numerically. 
In this letter we discuss the construction of the above mentioned 
basis for two dimensions. We
derive the system of quadratic equations and test the resulting 
Dirac operator in a dynamical simulation of the two-flavor Schwinger
model. 
\\
\\
\noindent
{\bf The Ginsparg-Wilson relation as a system of 
quadratic equations}
\vskip1mm
\noindent
We begin our construction with denoting the most general 2-d Dirac operator
on the lattice in the following form:
\begin{equation}
D_{x,y} \; = \; \sum_{\alpha = 0}^3 \Gamma_\alpha \; 
\sum_{p \in {\cal P}_{x,y}^\alpha} c_p^\alpha \; 
\prod_{l \in p} U_l \; .
\label{ansatz}
\end{equation}
Here $\Gamma_\alpha$ are the elements of the Clifford algebra, 
which in two dimensions has only 4 elements: $\Gamma_0 = $1\hspace*{-1mm}I,
$\Gamma_1 = \gamma_1$, $\Gamma_2 = \gamma_2$ and $\Gamma_3 = \gamma_5$.
The $\gamma_\mu$ are given by the Pauli matrices $\gamma_1 = \sigma_1,
\gamma_2 = \sigma_2$ and $\gamma_5 = \sigma_3$. To each generator 
$\Gamma_\alpha$ and to each pair of points $x,y$ on the lattice we assign
a set ${\cal P}_{x,y}^\alpha$ of paths $p$ connecting the two points. 
Each path is weighted with some complex weight $c_p^\alpha$ and
the construction is made gauge invariant by including the ordered product
of the gauge transporters $U_l$ ($\in$ U(N) or SU(N))
for all links $l$ of $p$. The action is 
then given by 
$S = \sum_{x,y} \overline{\psi}_x D_{x,y} \psi_y $,
where $x$ and $y$ run over all of the lattice. 

The next step is to impose on $S$ the symmetries which we want to 
maintain. Translation invariance requires the sets 
${\cal P}_{x,y}^\alpha$ to be invariant under simultaneous shifts of 
$x$ and $y$ and rotation invariance requires the terms to
be invariant when rotating $x$ and $y$ with respect to each other.
Parity ($(x_1,x_2) \rightarrow (x_1, -x_2)$) implies that for each
path $p$ (with coefficient $c_p^\alpha$)
we must include the parity-reflected copy with coefficient
$s_{parity}^\alpha \cdot c_p^\alpha$ where the signs $s_{parity}^\alpha$ are 
defined by $\gamma_2 \Gamma_\alpha \gamma_2 = s_{parity}^\alpha 
\cdot \Gamma_\alpha$. 

More interesting are the symmetries C and 
$\gamma_5$-hermiticity\footnote{$\; \gamma_5$-hermiticity is defined 
by requiring $\gamma_5 D \gamma_5 = D^\dagger$ for the lattice Dirac
operator $D$. Together with C-invariance 
this property essentially requires that
$\gamma_1$ and $\gamma_2$ come with derivative-like lattice terms, while
1\hspace*{-1mm}I and $\gamma_5$ should come with scalars and 
pseudo-scalars respectively. 
Thus $\gamma_5$-hermiticity can be viewed as a lattice 
remnant of the properties necessary for proving the CPT theorem in
Minkowski space.}. It is easy to see that both of them imply a relation
between the coefficient for a path $p$ and the coefficient of the inverse
path $p^{-1}$. Implementing both these symmetries firstly restricts all 
coefficients $c_p^\alpha$ to be either real or purely imaginary (for 
$\alpha = 0,1,2$ they are real and $c_p^3$ is purely imaginary). 
Secondly we find that the coefficient for a path $p$ and the coefficient
for its inverse $p^{-1}$ are equal up to a sign $s_{charge}^\alpha$ defined 
by $C \Gamma_\alpha C = s_{charge}^\alpha \cdot 
\Gamma_\alpha^T$, where $T$ 
denotes transposition and the charge conjugation matrix in our representation
is given by $C = \gamma_2$. 

\begin{figure}[htpb]
\centerline{
\epsfysize=4cm \epsfbox[ 0 0  543 210 ] {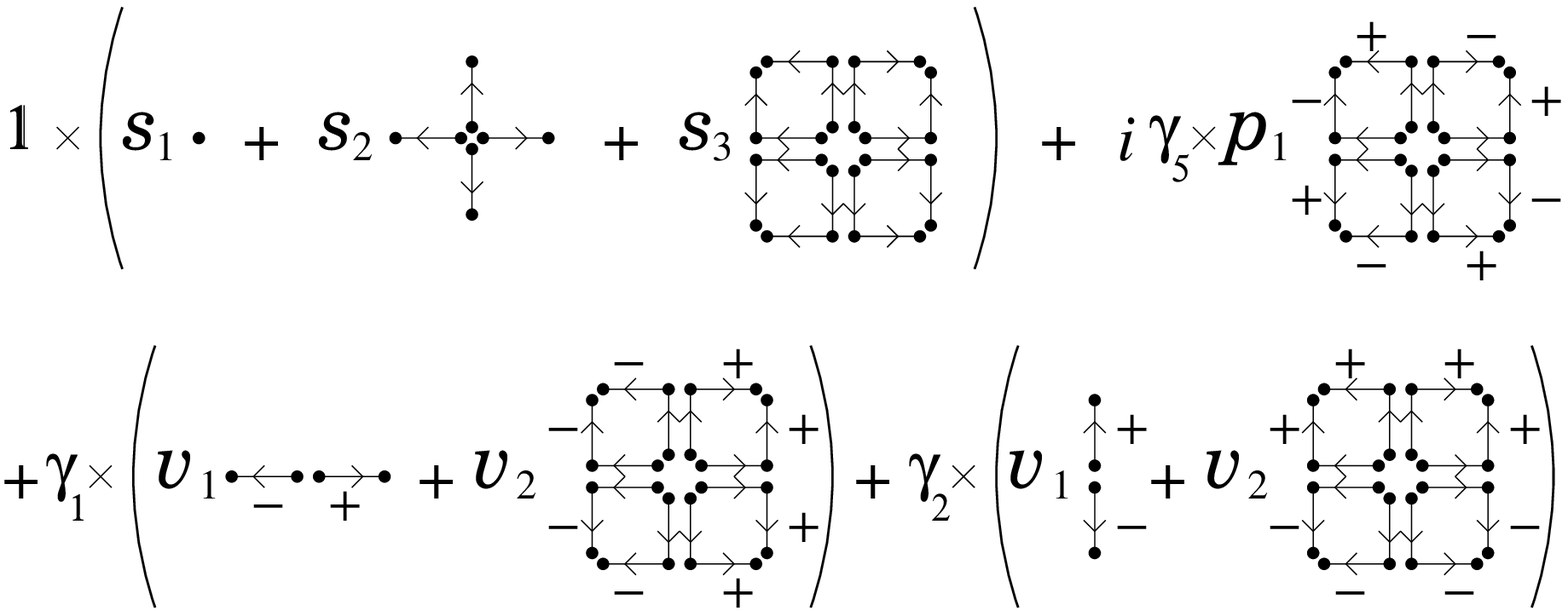}}
\caption{ {\sl Paths and coefficients for the Dirac operator}
D2.
\label{basicpaths}}
\end{figure}

When implementing all these symmetries we find that paths in our ansatz 
become grouped together in {\it diagrams} where, up to sign factors, 
all paths 
in a diagram come with the same coefficient. It is most convenient to 
represent the resulting Dirac operator using a pictorial notation. 
In Fig.~\ref{basicpaths} we display the leading diagrams in our expansion 
of the Dirac operator, i.e.~terms with paths on single plaquettes and
shorter than 2. The Dirac operator with this particular choice of 
terms will be referred to as D2 in the following.
\vskip1mm
\noindent
We have furnished the Clifford algebra element $\gamma_5$ with an additional
imaginary $i$, such  that the coefficients $s_1, s_2, s_3, p_1, v_1, v_2$
are real. For the generators $\gamma_5, \gamma_1$ and $\gamma_2$ we
assigned the relative signs for different paths in a diagram, according to
the rules discussed above. 
These signs appear next to the paths. For the generator 1\hspace*{-1mm}I
all paths come with plus signs and we omitted these in Fig.~1.
In order to be explicit about the
interpretation of our pictorial notation we give as an example
also the algebraic expression for the two $\gamma_1$-terms in D2:
\begin{eqnarray}  
& & v_1 \sum_{\mu = \pm 1} \mbox{sign}(\mu) \; U_\mu(x)
\; \delta_{x+\hat{\mu},y} \; 
\nonumber 
\\
& + & v_2 \sum_{{ \mu = \pm 1 \atop 
 \nu = \pm_2}} \mbox{sign}
(\mu) \; \left[ U_\mu(x) U_\nu(x + \hat{\mu}) +  U_\nu(x) 
U_\mu(x + \hat{\nu}) \right] \; \delta_{x+\hat{\mu}+ \hat{\nu},y} \; .  
\label{algebra}
\end{eqnarray} 
It is important to
remark, that the terms displayed in Fig.~1 are only the leading terms of
an infinite series of diagrams. Diagrams with arbitrarily 
long paths have to be included, 
since it is known, that there are no ultra-local
solutions of the Ginsparg-Wilson equation \cite{horvath}.

The next step in the derivation is to insert the diagrammatic expansion of 
$D$ into 
the Ginsparg-Wilson equation. To this purpose we define
\begin{equation}
E \; \equiv \; - D - \gamma_5 D \gamma_5 + \gamma_5 D \gamma_5 D \; ,
\label{giwi2}
\end{equation}
and finding a solution of the Ginsparg-Wilson equation corresponds to 
having $E = 0$.
The linear terms in (\ref{giwi2})
are easy to evaluate. The sandwich of $D$ 
with $\gamma_5$ leaves 
the terms with 1\hspace*{-1mm}I and $i \gamma_5$ in Fig.~1
unchanged, 
while the terms with $\gamma_1$ and $\gamma_2$ change their sign. These latter
terms thus cancel when adding $D$ and $\gamma_5 D \gamma_5$, while the
former terms obtain a factor of 2. 

Finally we have to compute the quadratic part $\gamma_5 D \gamma_5 D$.
Here we obtain products of diagrams, each of them with their respective 
Clifford algebra elements. 
The multiplication of these terms proceeds in two steps
(as can easily be
seen by writing out an example using the algebraic expression
(compare (\ref{algebra})) 
for our pictorial notation): In the first step 
the two elements of the Clifford algebra have to be multiplied. Since the 
Clifford algebra is closed the result is again a single element of the
algebra. In the second step the paths of the diagrams under consideration 
have to be glued together. In particular one takes a path appearing in 
$\gamma_5 D \gamma_5$ and continues it with a path appearing in $D$. When 
the latter path traces back some of the links of the former path, these 
links are removed from the diagram since the corresponding gauge 
transporters cancel each other. 

When now collecting all terms we obtain 
an expansion of the operator $E$ defined in (\ref{giwi2}) in terms of 
our diagrams\footnote{ A short comment is in order here: Obviously 
$E$ is a hermitian operator, while the diagrams for $\gamma_1$ and
$\gamma_2$ are anti-hermitian. These anti-hermitian terms, however, get 
transformed into their hermitian version when computing $E$ along the lines 
sketched above. The hermitian versions of the $\gamma_1$- and 
$\gamma_2$-diagrams differ from the original terms by a different 
assignment for the 
relative signs of the paths.}. 
The crucial step in our construction is to now view 
the diagrams in the expansion of $E$ as basis elements (it is easy to see
that they are linearly independent in an arbitrary background gauge field). 
Each of these basis elements is
multiplied with a quadratic polynomial in the coefficients 
$s_i, v_i, p_i$. In order to have $E = 0$, and thus a solution of the 
Ginsparg-Wilson equation, we have to find the zeros of all these 
polynomials simultaneously, i.e.~we have to solve 
a system of coupled quadratic equations. We have already remarked
that one needs infinitely
many diagrams (and coefficients $s_i, v_i, p_i$) for an exact solution 
of the Ginsparg-Wilson equation and thus the discussed set 
of equations is an infinite set 
of quadratic equations for the infinitely many coefficients $s_i, v_i, p_i$.
This infinite set of quadratic equations describes all possible solutions of
the Ginsparg-Wilson equation. Many interesting questions, such as how
different solutions are connected, can be formulated in terms of this
infinite set of equations. However, such questions will be pursued 
elsewhere \cite{future}. Here we now concentrate on solving a subset of
these equations and constructing approximate solutions of the 
Ginsparg-Wilson equation. 
\\
\\
\noindent
{\bf Construction of approximate solutions}
\vskip1mm
\noindent
From now on we will work with the truncated Dirac operator D2, represented
by the diagrams shown in Fig.~1. As already remarked above, such a
truncation
gives a meaningful approximation since, due to universality, we are only 
interested in Dirac operators where the coefficients of the paths decrease 
exponentially with the length of the paths. Of course this property has 
to be checked in the end. 

When deriving the expansion for $E$ along the lines of the last section,
now for a Dirac operator with only finitely many diagrams, we of course 
end up with only finitely many equations. For the case of D2 with its 
6 parameters $s_1, s_2, s_3, p_1, v_1, v_2$ we find all together 45 
quadratic equations. This system is overdetermined and we 
cannot satisfy all 45 equations with only 6 parameters. This property  
reflects the already mentioned fact, that there are no ultra-local
solutions of the Ginsparg-Wilson equation \cite{horvath}. If we were able 
to solve all 45 equations, $E$ would vanish exactly, and we would have found
a solution of the Ginsparg-Wilson equation with compact support. Thus,
with our 6 parameters we can only construct an approximate solution 
of the Ginsparg-Wilson equation. For this purpose we will select a 
dominating subset of our 45 equations and solve only this subset.

Before we come to this step, let's first discuss our system for trivial 
background field. In this case we can Fourier-transform the Dirac 
operator and the Fourier transform of any decent Dirac operator must
obey $\hat{D}(p) = i\not \mbox{\hspace{-1.4mm}} p + 
O(p^2)$, i.e.~the constant term 
must vanish and the linear term has to come with slope 1.
Fourier-transforming our D2 and implementing the above condition gives the 
following two equations for the coefficients:
\begin{eqnarray}
0 & = & s_1 \; + \; 4 s_2 \; + \; 8 s_3 \; ,
\label{m0}
\\
0 & = & 2 v_1 \; + \; 8 v_2 \; - \; 1 \; .
\label{c1}
\end{eqnarray}
These two equations must be implemented necessarily, such that we need 
only 4 of the equations from the expansion of $E$ in order to fix
the 6 parameters $s_1, s_2, s_3, p_1, v_1, v_2$. The basic criterion for
choosing the most dominant equations is clear: Equations corresponding to 
diagrams with shorter paths are more important, since the coefficients 
for diagrams with longer paths become exponentially small. Thus we 
first choose
the equations for the diagrams with factors $s_1, s_2$ and $s_3$ in Fig.~1. 
Furthermore, it can be shown to all orders, that there is no 
diagram in the expansion of $E$ which corresponds to the $v_1$ term 
of Fig.~1, since there exists no hermitian version of this diagram
(compare the second footnote). Thus the remaining terms are 
all of length 2 and we could choose as our
fourth equation either the
equation for the $p_1$ term or the equation for the 
hermitian version of the $v_2$ term. 
Here an interesting observation can be made: As we 
approach the continuum limit, the gauge transporters $U_l$ approach
1\hspace*{-1mm}I. Thus the weights from the gauge transporters
on the two paths leading to each point in 
the $p_1$ diagram of Fig.~1 approach each other as well. Since they 
come with opposite sign they cancel in the continuum limit. Thus this 
term takes care of itself in the continuum limit. The corresponding 
equation is not particularly important and we decided not to implement it
at this order. The same argument can be repeated for the hermitian version 
of the $v_2$ diagram. We instead use as our fourth equation 
one of the equations corresponding to a length-3 diagram in the scalar 
sector. In particular we choose the term with a maximum number of 
different paths. This term is unique and we end up with the following set of
equations,
\begin{eqnarray}
0 & = & s_1^2 - 2s_1 + 4 s_2^2 + 8 s_3^2 + 8 p_1^2 + 4v_1^2 + 16 v_2^2 \; ,
\\
0 & = & 2 s_1 s_2 - 2 s_2 + 4 s_2 s_3 + 4 v_1 v_2 \; ,
\\
0 & = & 2 s_1 s_3 - 2 s_3 + s_2^2 \; ,
\\
0 & = & s_2 s_3 - v_1 v_2 \; , 
\end{eqnarray}
which together with (\ref{m0}) and (\ref{c1}) are used to determine 
the coefficients. It is straightforward to find solutions using some 
numerical solver. Although we found several solutions, only one of
them provides a decent Dirac operator. However, 
since the unphysical solutions are very far away in 
the space of coefficients, identifying the interesting solution was 
not a problem. The coefficients for our solution 
are given in Table 1.
\begin{table}[h]
\begin{center}
\begin{tabular}{|c|c|c|c|c|c|}
\hline
 $s_1$ & $s_2$  & $s_3$ & $p_1$ & $v_1$ & $v_2$ \\
\hline
$ 1.26121 $ & $ -0.18471 $ & $ -0.06529 $ & $ -0.15745 $  
& $ 0.36940 $ & $ 0.03264 $ \\
\hline
\end{tabular}
\end{center}
\caption{ {\sl Values of the coefficients in the Dirac operator 
} D2.}
\end{table}
\noindent
It is obvious that the coefficients decrease quickly in size as the length 
of the paths increases. In order to properly show exponential decay one
would have to include more and longer diagrams. This would
also allow the Dirac operator to even better approximate a solution 
of the Ginsparg-Wilson equation. However, already with 
our relatively modest expense in additional terms we find very good
approximate chirality. This property will be explored numerically
in the next section. 
\\
\\
\noindent
{\bf Results from the numerical simulation with D2}
\vskip1mm
\noindent
We now choose our gauge group to be U(1). We first show two plots 
of the complex eigenvalues for our operator D2 in gauge field 
configurations from quenched ensembles at $\beta = 6$ and $\beta = 4$.
\begin{figure}[htpb]
\centerline{\hspace*{-3mm}
\epsfig{file=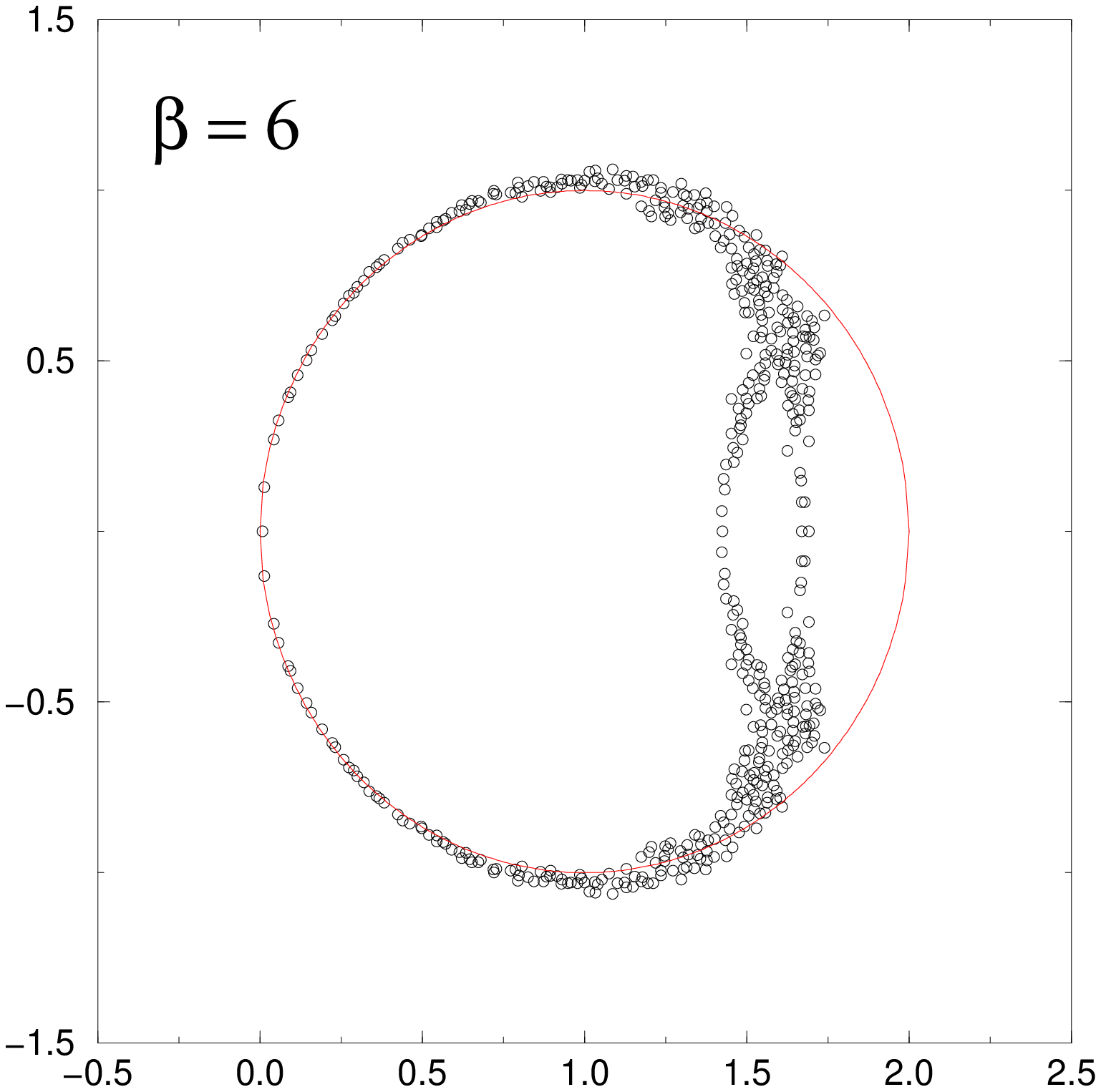,width=6cm,height=5.5cm}
\epsfig{file=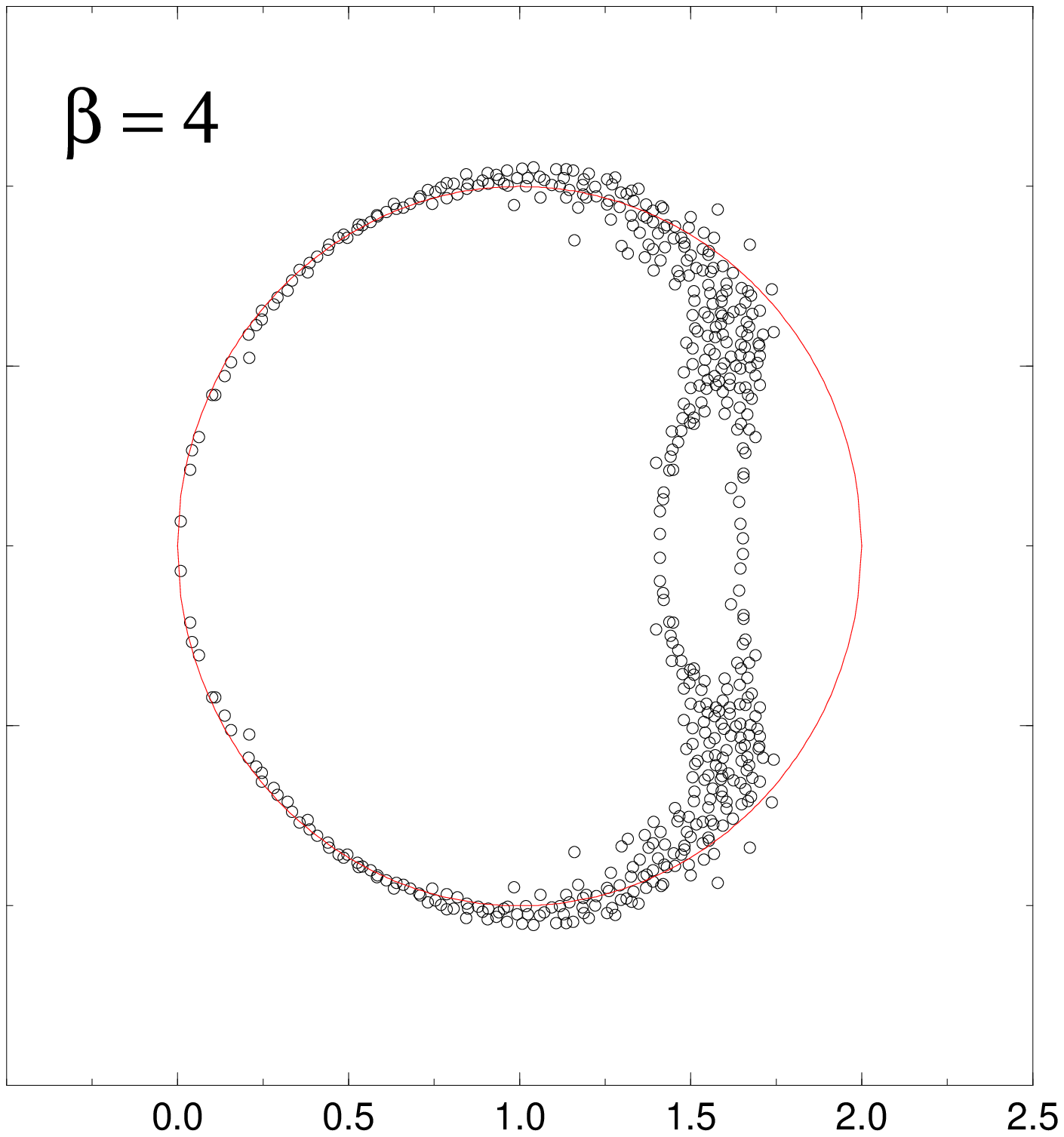,width=5.4cm,height=5.4cm}
}
\caption{ {\sl Plots of spectra of} D2 {\sl in the complex plane for
quenched gauge field configurations at $\beta = 6$ (left-hand side) 
and $\beta = 4$. The circle indicates the spectrum for exact solutions
of the Ginsparg-Wilson equation.}  
\label{spectra}}
\end{figure}
\vskip5mm
\noindent
It is easy to show, that (1) together with $\gamma_5$-hermiticity 
implies that the eigenvalues of an exact solution of the 
Ginsparg-Wilson equation lie on 
a circle of radius 1 around 1 in the complex
plane. We find that the physical sector of the spectrum of D2 is very
close to this limiting circle, while the doubler branches deviate 
considerably. This is not particularly disturbing, since the doublers 
decouple from the physical spectrum when driving the system to the
continuum limit. 
Most importantly, however, we find that the physical 
branch intersects the real axis very close to the origin, rendering 
essentially massless fermions. This holds for both the $\beta = 6$ and 
the $\beta = 4$ ensemble. We furthermore find, that also the fluctuations 
of the eigenvalues at the origin are much smaller than the fluctuations
for e.g.~the standard Wilson operator with a fine tuned mass term. This
leads to error bars for the $\pi$-mass (compare below) which are one 
order of magnitude smaller than with the standard Wilson action 
\cite{GaHiLa99}.  

In order to further test the chiral properties of D2 we performed a 
simulation of the massless 2-flavor Schwinger model with dynamical 
fermions. In the continuum this model 
can be solved explicitly and it contains an iso-triplet of massless 
$\pi$-particles and a massive iso-singlet 
$\eta$ (for the Schwinger model with flavor 
see e.g.~\cite{schwinger}). When putting the
model on the lattice a violation of the chiral symmetry will 
lead to a non-vanishing
mass for the $\pi$. The smallness of $m_\pi$ thus is a measure for 
the quality of any approximation to the Ginsparg-Wilson equation. 
In our simulation we compute the masses and the dispersion relations
for both the $\pi$ and the $\eta$. 

We compare our results to the numerical results for the two known
exact solutions of the Ginsparg-Wilson equation, i.e.~the overlap 
Dirac operator and the perfect action. The data for these cases were taken 
from \cite{fix2db}. 
Furthermore we compare our results also to the results for 
an alternative approximate 
solution of the Ginsparg-Wilson equation presented in \cite{BiHi99}.
In particular we compare with the truncated perfect operator augmented
with an additional Clover term (TP+C). We remark, that also finite size 
effects can contribute to $m_\pi$. However, we estimated these effects to be 
smaller than our error bars. Furthermore, all data presented in Table 2 
were computed for the same lattice size ($16^2$) such that for our direct 
comparison of lattice results finite size effects are irrelevant.  
All data shown in Table 2 have the 
same statistics ($10^4$ measurements).
\begin{table}[htpb]
\begin{center}
\begin{tabular}{|c|c|c|c|c|c|}
\hline
 & D2 & TP+C & perfect & overlap & continuum\\
\hline
$ m_\pi $ & 0.0340(13) & 0.1304(27) & 0.0088(5) & 0.0038(23) & 0.0 \\
\hline
$ m_\eta$ & 0.334(27) & 0.347(10)  & 0.351(10)  & 0.338(13) & 0.325730 \\
\hline
\end{tabular}
\end{center}
\caption{ {\sl $\pi$ and $\eta$ mass at $\beta = 6$ for 
various lattice Dirac operators as well as their values in the 
continuum. }}
\end{table}
\vskip1mm
\noindent
In Table 2 we show, for $\beta = 6$, 
the masses for these different approaches and also 
give the continuum values. For the $\pi$-mass we find a value of
0.0340 which is about four times as large as the mass obtained with 
the perfect action and about ten times as large as the mass from overlap.
These Dirac 
operators are on the other hand much more expensive to simulate and
still give a non-vanishing $m_\pi$ due to cut-offs or numerical 
errors. When 
comparing to the other approximate solution, the TP+C operator, we find
that we have reduced the $\pi$-mass by a factor of 4, while the cost of the 
numerical simulation remained the same. We do even better for the 
$\eta$-mass where our data provides the best approximation of the
continuum result. 

\begin{figure}[htpb]
\centerline{
\epsfysize=5.7cm \epsfbox[ 30 23 522 438 ] {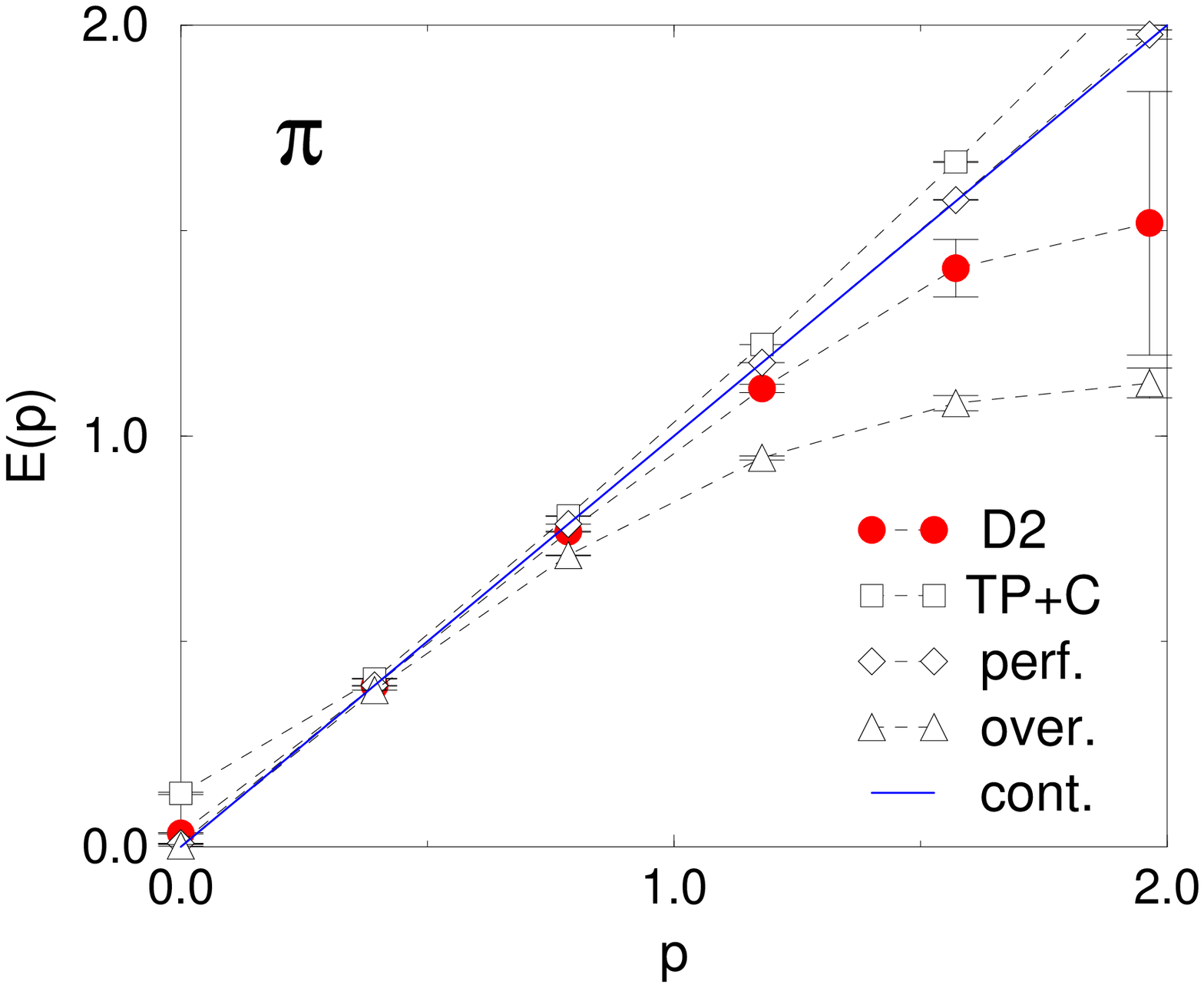}
\epsfysize=5.7cm \epsfbox[ 104 23 522 438 ] {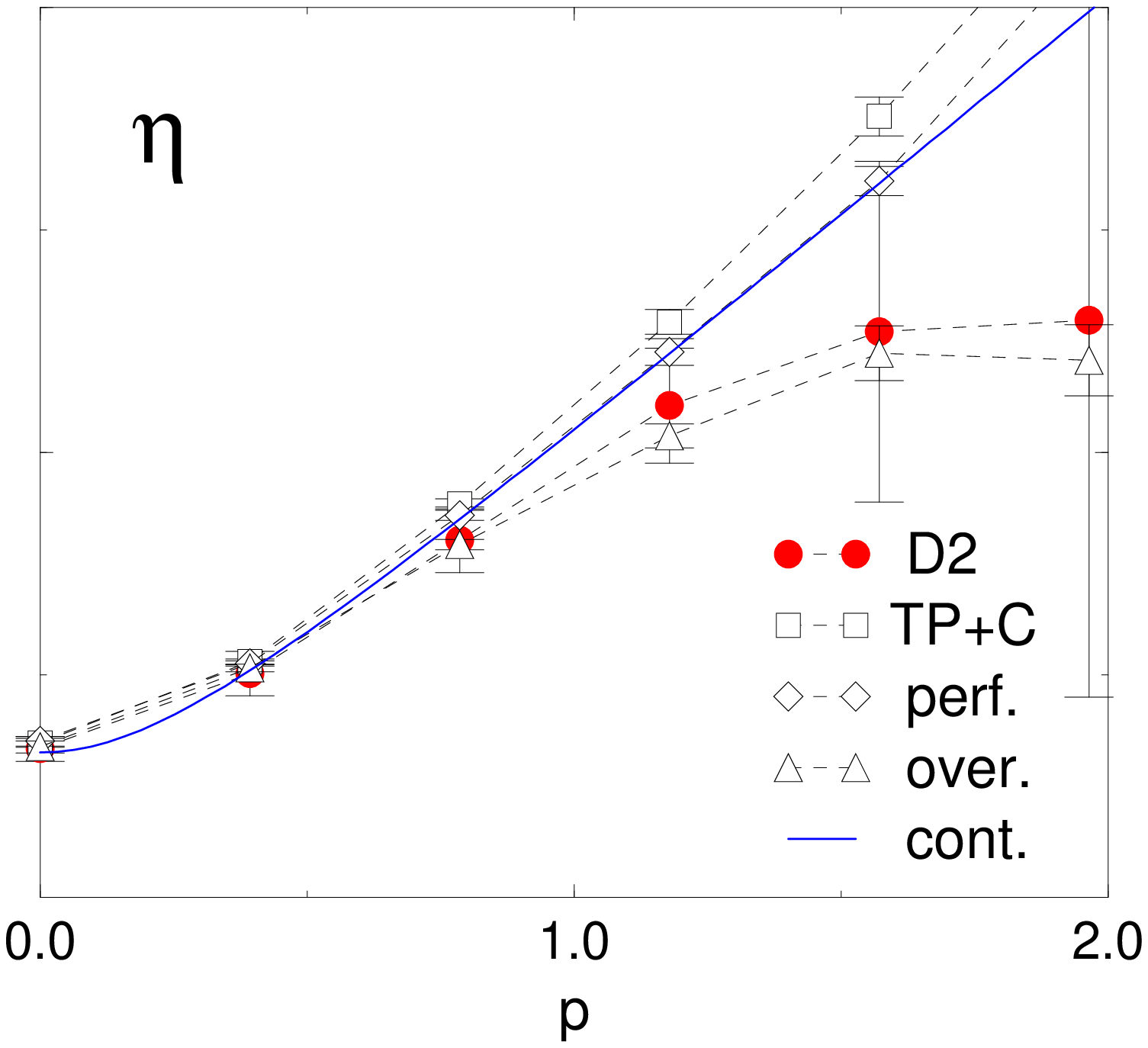}}
\caption{ {\sl Dispersion relations for $\pi$ (left-hand side)
and $\eta$. 
The filled circles are the numerical data for our operator} D2, {\sl the 
triangles represent the overlap operator, diamonds the perfect
action and we use squares for} TP+C. {\sl The symbols are 
connected with dashed lines to guide the eye. The full line is the 
continuum result.}  
\label{fullplot}}
\end{figure}

Let's now compare the dispersion relations. 
In Fig.~\ref{fullplot} we show the $\pi$ and $\eta$ dispersion 
relations of
our operator D2 (filled circles) 
and compare it with the dispersion relations of
the other lattice approaches (symbols) 
as well as with the continuum result (full curve). 
\noindent
For the $\pi$ dispersion relation we find that the our operator
leads to a nice linear behavior up to $p \sim 1$. For larger 
values of $p$ the dispersion curve begins to deviate from the
continuum value, although not as strongly as the overlap curve.
Only the perfect action manages to obtain values close to the 
continuum result throughout. For larger momenta also the TP+C operator
is relatively close to the continuum curve, however, for 
$p \rightarrow 0$ the violations of chirality are obvious. 

For the
$\eta$ the picture is similar: Only the perfect action has a 
perfect dispersion relation. The other actions lead to dispersion 
relations which for increasing momenta deviate more or less from the
continuum result. Again we find that our D2 curve lies between the 
continuum curve and the data for the overlap operator. 

\newpage
\noindent
{\bf Summary}
\vskip1mm
\noindent 
In this letter we have explored a new approach to Ginsparg-Wilson fermions
and tested the new method in 2-d. 
To keep the paper self consistent the construction of the Dirac 
operator was also given in 2-d, but the generalization to 4-d is 
straightforward and will be presented elsewhere \cite{future}. 

The central idea is to construct an expansion of the Dirac operator
in a suitable basis and to rewrite the Ginsparg-Wilson equation 
to a system of coupled quadratic equations for the coefficients.
The basis elements are realized by diagrams of paths on the lattice,
with the paths constituting a diagram being related by symmetries
(rotations, C, P and $\gamma_5$-hermiticity). Although the expansion 
in principle has to contain infinitely many terms, the method
comes with a natural cutoff: Only Dirac operators where the coefficients 
decrease exponentially with increasing length of the paths are of
interest. Thus we can work with a finite number of terms, rendering
a finite system of coupled quadratic equations. The dominating equations
of the system can be solved and used to construct approximate 
solutions of the Ginsparg-Wilson equation.

We implemented this procedure and constructed the operator 
D2 which in addition to 
the terms appearing in the standard Wilson action only requires L-shaped
terms of length 2. These terms imply only a moderate increase in the
cost of a numerical simulation but drastically improve the chiral
properties of the action. This was established in a numerical 
simulation of the 2-flavor Schwinger model. 

We are currently 
implementing the new approach in 4-d, where the construction proceeds 
essentially along the lines presented here. We expect that also in 4-d
a considerable improvement of the chiral properties can be achieved with 
only a moderate increase of the cost of a numerical simulation.
\\
\\
{\sl Acknowledgement:} The authors thank Thomas Lippert 
and Uwe-Jens Wiese for insightful discussions and Christian Lang for 
important advice and checks of our equations.

\end{document}